%% file: main.tex
\documentclass[10pt]{article} 
\usepackage[accepted]{rlc}

\usepackage{amssymb}            
\usepackage{mathtools}          
\usepackage{mathrsfs}           
\usepackage{graphicx}           
\usepackage{subcaption}         
\usepackage[space]{grffile}     
\usepackage{url}                

\usepackage{bm}
\usepackage{booktabs}
\usepackage{algorithm}
\usepackage{algorithmicx}
\usepackage[noend]{algpseudocode}
\usepackage{enumitem}

\input{collections/macros}

\title{Towards General Negotiation Strategies with\\End-to-End Reinforcement Learning}

\author{Bram M. Renting\\
    b.m.renting@liacs.leidenuniv.nl\\
    Leiden University\\
    Delft University of Technology
    \And
    Thomas M. Moerland\\
    t.m.moerland@liacs.leidenuniv.nl\\
    Leiden University
    \AND
    Holger H. Hoos\\
    hh@cs.rwth-aachen.de\\
    RWTH Aachen University\\
    Leiden University
    \And
    Catholijn M. Jonker\\
    c.m.jonker@tudelft.nl\\
    Delft University of Technology\\
    Leiden University
}

\begin{document}

\maketitle
\begin{abstract}
\input{sections/0_abstract}
\end{abstract}

\section{Introduction}\label{sec:introduction}
\input{sections/1_introduction}

\section{Related Work}
\input{sections/2_related_work}

\section{Methods}
\input{sections/3_methods}

\section{Emperical Evaluation}
\input{sections/4_emperical_evaluation}

\section{Conclusion}\label{sec:conclusion}
\input{sections/5_conclusion}

\subsubsection*{Broader Impact Statement}\label{sec:broaderImpact}
It is often envisioned that negotiating agents will represent humans or other entities in a future where AI is more integrated into society. Having access to more capable negotiation agents could increase inequalities in such societies, especially if the development of such agents is a highly skilled endeavour. Removing the human aspect in negotiation might also lead to more self-centred behaviour. We should ensure that we design for fairness and cooperative behaviour in such systems.

\subsubsection*{Acknowledgments}\label{sec:ack}
This research was (partly) funded by the \href{https://hybrid-intelligence-centre.nl}{Hybrid Intelligence Center}, a 10-year programme funded by the Dutch Ministry of Education, Culture and Science through the Netherlands Organisation for Scientific Research, grant number  024.004.022 and by EU H2020 ICT48 project``Humane AI Net'' under contract \(\# \) 952026. This research was also partially supported by TAILOR, a project funded by the EU Horizon 2020 research and innovation programme under GA No 952215.


\bibliography{references}
\bibliographystyle{rlc}

\appendix
\section{PPO training hyperparameters}\label{app:hp}
\input{sections/A_hyperparameters}

\end{document}

%% file: collections/macros.tex
\DeclareMathOperator*{\argmax}{arg\,max}

{}
{}

%% file: sections/0_abstract.tex
The research field of automated negotiation has a long history of designing agents that can negotiate with other agents. Such negotiation strategies are traditionally based on manual design and heuristics. More recently, reinforcement learning approaches have also been used to train agents to negotiate. However, negotiation problems are diverse, causing observation and action dimensions to change, which cannot be handled by default linear policy networks. Previous work on this topic has circumvented this issue either by fixing the negotiation problem, causing policies to be non-transferable between negotiation problems or by abstracting the observations and actions into fixed-size representations, causing loss of information and expressiveness due to feature design. We developed an end-to-end reinforcement learning method for diverse negotiation problems by representing observations and actions as a graph and applying graph neural networks in the policy. With empirical evaluations, we show that our method is effective and that we can learn to negotiate with other agents on never-before-seen negotiation problems. Our result opens up new opportunities for reinforcement learning in negotiation agents.

%% file: sections/1_introduction.tex
In multi-agent systems, agents sometimes must coordinate actions to improve payoff or even obtain payoff in the first place (e.g., surveying drone swarms or transporting goods using multiple robots). In such scenarios, communication between agents can improve insight into other agents' intentions and behaviour, leading to better coordination between agents and thus improving payoff. When agents have individual preferences besides a shared common goal, also known as mixed-motive or general sum games, communication can become more complex, as this introduces an incentive to deceive~\citep{dafoe_open_2020}.

A special case of communication in mixed-motive multi-agent systems is negotiation, which allows for finding and agreeing on mutually beneficial coordinated actions before performing them. Negotiation plays a central role in many present and future real-world applications, such as traffic light coordination, calendar scheduling, or balancing energy demand and production in local power grids, but also in games, such as Diplomacy or Werewolves. Automated Negotiation is a long-standing research field that has focussed on designing agents that can negotiate~\citep{smith_contract_1980,rosenschein_rational_1986,sycara_resolving_1988,tawfik_jelassi_negotiation_1989,klein_conflict_1989,robinson_negotiation_1990}.

Traditionally, many negotiating agents were manually designed algorithms based on heuristics, which is still a commonly seen approach in recent editions of the Automated Negotiation Agents Competition (ANAC)~\citep{aydogan_13th_2023}. However, manually designing such negotiation strategies is time-consuming and results in highly specialised and fixed negotiation strategies that often do not generalise well over a broad set of negotiation settings. In later work, optimisation methods were used to optimise the parameters of negotiation strategies using evolutionary algorithms~\citep{eymann_co-evolution_2001,dworman_bargaining_1996,lau_evolutionary_2006}, or algorithm configuration techniques~\citep{renting_automated_2020}. Such approaches allow negotiation strategies to be more easily adaptable to different negotiation problems but still require partial manual design to obtain a parameterized negotiation strategy, making them cumbersome and limiting their generalizability.

With the advent of Reinforcement Learning (RL)~\citep{sutton_reinforcement_2018}, there have been attempts at using RL-based methods for creating negotiation agents~\citep{bakker_rlboa_2019}. There is, however, still an open challenge. In automated negotiation, it is common for agents to deal with various negotiation problems that would cause differently sized observation and action vectors for default linear layer-based RL policies. Up until now, this issue has been dealt with by either abstracting the observations and actions into a fixed-length vector (see, e.g., \citet{bakker_rlboa_2019}) or by fixing the negotiation problem, such that the observation and action space remain identical (see, e.g., \citet{higa_reward-based_2023}). The first approach causes information loss due to feature design, and the latter renders the obtained policy non-transferable to other negotiation problems.

We set out on the idea that a more general RL-based negotiation strategy capable of dealing with various negotiation problems is achievable and that such a strategy can be learned using end-to-end reinforcement learning without using state abstractions. Developing such an RL negotiation strategy would open up new avenues for RL in automated negotiation as policies are easily extendable. End-to-end methods are also able to learn complex relations between observations and actions, minimizing the risk of information loss that is often imposed by (partially) manual design strategies.

To this extent, we designed a graph-based representation of a negotiation problem. We applied graph neural networks in the RL policy to deal with the changing dimensions of both the observation and action space. To the best of our knowledge, graph-based policy networks have not been used before to handle changing action spaces, except by \citet{yang_transfer_2024}, who independently proposed a similar approach to another problem. We show that our method shows similar performance to a recent end-to-end RL-based method designed to deal only with a fixed negotiation problem. More importantly, we show that our end-to-end method can successfully learn to negotiate with other agents and that the obtained policy still performs on unseen, randomly generated negotiation problems.

%% file: sections/2_related_work.tex
\citet{bakker_rlboa_2019} applied RL to decide what utility to demand in the next offer. They abstracted the state to utility values of the last few offers and time towards the deadline. Translating utility to an offer, estimating opponent utility, and deciding when to accept were done without RL. \citet{bagga_deep_2022} also abstracted the state into a fixed representation with utility statistics of historical offers. They used an RL policy to decide whether to accept and a separate policy that outputs offers based on a non-RL opponent utility estimation model.

\citet{sengupta_autonomous_2021} encoded the state into a fixed length of past utility values. The action is the utility offer target, translated to an actual offer through an exhaustive search of the outcome space. They trained a portfolio of policies and tried to select effective counterstrategies by classifying the opponent type. \citet{li_combining_2023} also build a portfolio of RL-based negotiation strategies by incrementally training best responses based on the Nash bargaining solution. During evaluation, their method searches for the best response in an effort to improve cooperativity. They only applied their method to fixed negotiation problems.
 
Another line of research on negotiation agents includes natural language. An environment for this was developed by \citet{lewis_deal_2017}. \citet{kwon_targeted_2021} used this environment and applied a combination of RL, supervised learning, and expert annotations (based on a dataset) to iteratively train two agents through self-play. The negotiation problems considered are fixed, except for the preferences. 

\cite{takahashi_venas_2022} and \citet{higa_reward-based_2023} are closest to our work, as they also train an end-to-end RL method for negotiation games. Their approach does not use state abstractions and linearly maps the negotiation problem and actions in a policy. The policy obtained can only be used for a fixed problem. They also trained and tested only against single opponents.

%% file: sections/3_methods.tex
We formulate the negotiation game as a turn-based Partially Observable Stochastic Game (POSG), a partially observable extension of a stochastic game~\citep{shapley_stochastic_1953}. We model the game as a tuple $\mathcal{M} = \langle \mathcal{I}, \mathcal{S}, \mathcal{O}_i, \mathcal{A}_i, \mathcal{T}, \Omega_i, \mathcal{R}_i \rangle$, where $\mathcal{I} = \{1, \cdots, n\}$ denotes the set of agents, 
$\mathcal{S}$ the set of states,
$\mathcal{O}_i$ the set of possible observations for agent $i$, and
$\mathcal{A}_i$ the set of actions for agent $i$. For convenience, we write $\mathcal{A} = \mathcal{A}_i$, as we consider a turn-based game where only single agents take actions. Furthermore, 
$\mathcal{T} : \mathcal{S} \times \mathcal{A} \mapsto p(\mathcal{S})$ denotes the transition function,
$\Omega_i : \mathcal{S} \times \mathcal{A} \mapsto p(\mathcal{O}_i)$ the observation function for agent $i$, and $\mathcal{R}_i : \mathcal{S} \times \mathcal{A} \mapsto \mathbb{R}$ the reward function for agent $i$. 

The game starts in a particular state $s$. Then, at timestep $t$, an agent $i$ selects an action $a_{t,i}$ independently of other agents. Based on this action, the state of the POSG changes according to $s_{t+1} \sim \mathcal{T}(s_{t+1}|s_t,a_t)$. Subsequently, each agent receives its own observation $o_{t,i} \sim \Omega_i(o_{t,i}|s_t,a_t)$ and associated reward $r_{t,i} \sim \mathcal{R}_i(r_{t,i}|s_t,a_t)$.

Each agent $i$ selects actions according to its own policy $\pi_i: \mathcal{O}_i \times \mathcal{O}_i \times \cdots \to p(\mathcal{A})$. At timestep $t$, agent $i$ samples an action $a_{t} \sim \pi_i(a_{t} | o_{t,i},o_{t-1,i}, \cdots)$. Note that we can vary the length of the historical observations by which we condition the policy for each agent. The more history we include, the more we can overcome partial observability.

Our goal is to find a policy $\pi_i$ for agent $i$ that maximizes its cumulative expected return:
\begin{equation}
\pi_i^\star \in \argmax_{\pi_i} \mathbb{E}_{\pi, \mathcal{T}} \left[ \sum_{k=0}^H \mathcal{R}_i(s_{t+k},a_{t+k}) \right],
\end{equation}

where $H$ denotes the horizon of the POSG (the number of rounds we select an action). Crucially, the performance of a particular policy $\pi_i$ depends on the other agents' policies.

\subsection{Negotiation Game}
A negotiation game consists of a set of agents and a problem to negotiate over. This work only considers bilateral negotiation games with two agents. The negotiation problem, also known as a negotiation domain, generally consists of a set of objectives (or issues) $B = \{1,\cdots,m\}$ with an associated set of values $V_b$ to choose from. Value sets can be continuous, integer, or discrete, but we focus solely on discrete value sets in this work, which is the most general type, as continuous values can also be discretised. For each of the objectives $b \in B$, both agents must agree upon a value $v_b \in V_b$. The total outcome space is the Cartesian product of all the value sets $\Omega = V_1 \times \cdots \times V_{m}$ with a single outcome being $\omega = \langle v_1, \cdots, v_m \rangle$.

Both agents have preferences over the outcome space expressed through a utility function $u: \Omega \mapsto [0,1]$ that is private information. Here, 1 is their best possible outcome, and 0 is their worst. This paper only considers additive utility functions as shown in \autoref{eq:utility}. Here, weights are assigned to all values and objectives through weight functions $w: B \mapsto [0,1]$ and $w_b: V_b \mapsto [0,1]$ such that $\sum_{b \in B} w(b) = 1$, $\max_{v_b \in V_b} w_b(v_b) = 1$, and $\min_{v_b \in V_b} w_b(v_b) = 0$.

\begin{equation}\label{eq:utility}
    u(\omega) = \sum_{b \in B} w(b) \cdot w_b(v_b)
\end{equation}

\subsubsection{Protocol}
The negotiation follows the commonly used Alternating Offers Protocol~\citep{rubinstein_perfect_1982}, where agents take turns. During its turn, an agent can make a (counter) offer or accept the opponent's offer. A deadline is imposed to prevent the agents from negotiating indefinitely. Failure to reach an agreement before the deadline results in 0 payoff. When an agreement is reached, both agents obtain the payoff defined by their utility function.

\subsection{PPO}
We will use reinforcement learning to optimize the policy $\pi_i$ of our own agent $i$ in the negotiation problem. For simplicity, we will drop the subscript $i$ and simply write $\pi$ for the policy of our own agent. We also simplify by writing $o$ instead of $\langle o_{t,i},o_{t-1,i}, \cdots \rangle$. To optimize this policy, we use Proximal Policy Optimisation (PPO)~\citep{schulman_proximal_2017} due to its empirical success and stability. 

At each update iteration $k$, PPO optimises $\pi$ relative to the last policy $\pi_{k}$ by maximising the PPO clip objective:

\begin{equation}\label{eq:ppo}
    \pi_{k+1} \in \argmax_{\pi} \mathbb{E}_{o,a \sim \pi_{k}} \left[ \min \left( \frac{\pi(a|o)}{\pi_k(a|o)} A_{\pi_k}(o,a),\ \text{clip}\left( \frac{\pi(a|o)}{\pi_k(a|o)}, 1 \pm \epsilon \right) A_{\pi_k}(o,a) \right) \right]
\end{equation}

where $\epsilon$ denotes a clip parameter, and $A_{\pi}(a,o)$ denotes the advantage function of taking action $a$ in observation $o$~\citep{sutton_reinforcement_2018}. The ratio gets clipped to ensure that the new policy does not change too quickly from the policy at the previous step. Our PPO implementation is based on the CleanRL repository~\citep{huang_cleanrl_2022}.

\subsection{Graph Neural Networks}
We aim to learn to negotiate across randomly generated problems where the number of objectives and values differ. This forces us to design a policy/value network where the shape and number of parameters are independent of the number of objectives and values. Networks of linear layers, often the default in RL, do not fit this criterion, as they require fixed input dimensions. We chose to represent the input of the policy network as a graph and make use of Graph Neural Networks (GNN) to deal with the changing size of the input space, more specifically, Graph Attention Networks (GAT)~\citep{velickovic_graph_2018}.

The input graph contains nodes that have node features. A layer of GNN encodes the features $x_u$ of node $u$ into a hidden representation $h_u$ based on the features of the set of neighbour nodes $\mathcal{N}_u$ and on its own features. The specific case of GATs is defined in \autoref{eq:gat}. Here, neighbour features are encoded by a linear layer $\psi$ and then weighted summed through a learned attention coefficient $a(x_u, x_v)$. The weighted sum is concatenated with $x_u$ and passed through another linear layer $\phi$ to obtain the embedding of the node $h_u$.

\begin{equation}\label{eq:gat}
    h_u = \phi \left( x_u, \sum_{v \in \mathcal{N}_u} a(x_u, x_v) \cdot \psi (x_v) \right)
\end{equation}

\subsection{Implementation}
At each timestep, the agent receives observations that are the actions of the opponent in the negotiation game. Based on these observations, the agent must select an action. The action space combines multiple discrete actions: the accept action and an action per objective to select one of the values in that objective as an offer. If the policy outputs a positive accept, then the offer action becomes irrelevant as the negotiation will be ended.

\begin{figure}
    \centering
    \includegraphics[width=\textwidth]{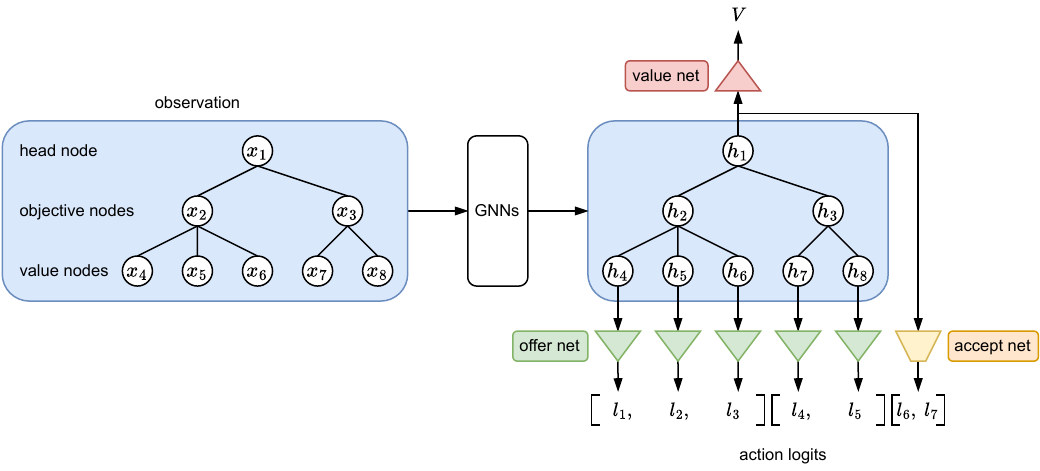}
    \caption{Overview of our designed policy network based on GNNs. Observations are encoded in a graph representation (left) and passed through GNNs. Action distribution logits and state-value are obtained by passing the learned representation of the head node and value nodes through linear layers.}
    \label{fig:state_graph}
\end{figure}

A negotiation problem has objectives $B$ and a set of values to decide on per objective $V_b$. We represent the structure of objectives and values as a graph and encode the history of observations $\langle o_{t,i},o_{t-1,i}, \cdots \rangle$ of a negotiation game in this structure to a single observation $o$ (see the left side of \autoref{fig:state_graph}). Each objective and value is represented by a node, where value nodes are connected to the objective node to which they belong. An additional head node is added that is connected to all objective nodes. The node features of each node are:
\begin{itemize}[nosep]
    \item 5 features for each value node: the weight $w_b(v_b)$ of the value, a binary value to indicate the opponent's last offer, a binary value to indicate the last offer of the agent itself, the fraction of times this value was offered by the opponent, and the fraction of times this value was offered by itself. Note that it might have been better to implement a recurrent network to condition the policy on the full history of offers instead of summary statistics. However, the added computational complexity would have rendered this work much more difficult. Our approach enables efficient learning, but future work should explore the use of the raw history of offers.
    \item 2 features for each objective node: the number of values in the value set of this objective $|V_b|$, and the weight of this objective $w(b)$.
    \item 2 features for the head node: the number of objectives $|B|$, and the progress towards the deadline scaled between 0 and 1.
\end{itemize}

As illustrated in \autoref{fig:state_graph}, we apply GAT layers to the observation graph to propagate information through the graph and embed the node features (\autoref{eq:gat}). The size of the representation is a hyperparameter. We then take the representation of the head node and pass it to a linear layer that predicts the state value $V$. The head representation is also passed through a linear layer to obtain the two accept action logits. Finally, we take the representation of every value node and apply a single linear layer to obtain the offer action logits. These logits are concatenated per action and used to create the probability distribution over the action space. As we use the same linear layer for all value nodes, the number of output logits is independent of the number of parameters in the policy, thus satisfying our requirement. We also note that although the size of the outcome space suffers heavily from the curse of dimensionality when the number of objectives increases, our approach does not. Our code implementation can be found on GitHub\footnote{\url{https://github.com/brenting/RL-negotiation/tree/RLC-2024}}.

%% file: sections/4_emperical_evaluation.tex
To train our agent, we need both negotiation problems and opponents to negotiate against. The negotiation problems will be randomly generated with an outcome space size $|\Omega|$ between 200 and 1000. As opponents, we use baseline agents and agents developed for the 2022 edition of the Automated Negotiation Agents Competition (ANAC). The baseline agents are simple negotiation strategies often used within automated negotiation to evaluate and analyse new agents. We provide a description of the opponents in \autoref{tab:opp}. All agents were originally developed for the GENIUS negotiation software platform~\citep{lin_genius_2014}.

\begin{table}
    \centering
    \begin{tabular}{lll}
        \toprule
        \textbf{Name} & \textbf{Type} & \textbf{Description} \\
        \midrule
        BoulwareAgent & Time-dependent & Utility target decreases concave with time \\
        ConcederAgent & Time-dependent & Utility target decreases convex with time \\
        LinearAgent & Time-dependent & Utility target decreases linearly with time \\
        RandomAgent & Random & Makes random offers, accepts any utility > $0.6$ \\
        \bottomrule
    \end{tabular}
    \caption{Description of baseline negotiation agents used for benchmarking.}
    \label{tab:opp}
\end{table}

We set a negotiation deadline of 40 rounds. An opponent is randomly selected during the rollout phase, and a negotiation problem is randomly generated. The policy is then used to negotiate until the episode ends, either by finding an agreement or reaching the deadline. The episode is added to the experience batch, which is repeated until the experience batch is full. We apply 4 layers of GATs with a hidden representation size of 256. A complete overview of the hyperparameter settings can be found in \autoref{app:hp}.

\subsection{Fixed Negotiation Problem}
As a first experiment, we compare our method to a recent end-to-end RL method by \citet{higa_reward-based_2023} that can only be used on a fixed negotiation problem. Their method was originally only trained and evaluated against single opponents. We chose to train the agent against the set of baseline players instead, as we consider that a more realistic scenario. The baseline agents show relatively similar behaviour, making this an acceptable increase in difficulty.

\begin{figure}
    \centering
    \includegraphics{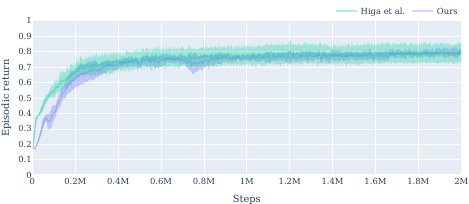}
    \caption{Mean and 99\% confidence interval of episodic return during training based on results from 10 random seeds. The results of the policy designed by \citet{higa_reward-based_2023} and our policy are plotted.}
    \label{fig:training_fixed}
\end{figure}

\begin{figure}
    \centering
     \begin{subfigure}{0.45\textwidth}
         \centering
         \includegraphics{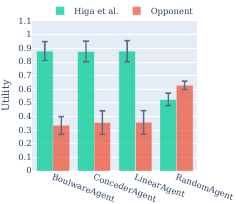}
     \end{subfigure}
     \begin{subfigure}{0.45\textwidth}
         \centering
         \includegraphics{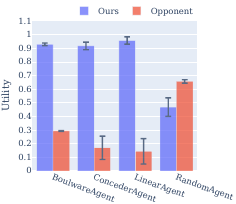}
     \end{subfigure}
    \caption{Evaluation results of the policy designed by \citet{higa_reward-based_2023} and our GNN-based policy. Results are obtained by evaluating each trained policy for 1000 negotiation games against the set of baseline agents. Mean and 99\% confidence interval are plotted based on 10 training iterations.}
    \label{fig:fixed}
\end{figure}

We generated a single negotiation problem and trained a reproduction of their and our own method for $2\,000\,000$ timesteps on 10 different seeds. The training curve is illustrated in \autoref{fig:training_fixed}, where we plot both the mean of the episodic return and the 99\% confidence interval based on the results from 10 training sessions. Every obtained policy is evaluated in 1000 negotiation games against every opponent on this fixed negotiation problem. We report the average obtained utility of the trained policy and the opponent, including a confidence interval based on the 10 evaluation runs in \autoref{fig:fixed}.

We can see in \autoref{fig:fixed} that our method performs similarly to the method proposed by \citet{higa_reward-based_2023}. This result is mostly a sanity check that our method can successfully learn to negotiate in a relatively simple setup despite being more complex and broadly usable.

\subsection{Random Negotiation Problems}
We now evaluate the performance of our end-to-end method on randomly generated negotiation problems. Negotiation problems will continuously change during both training and evaluation.

\begin{figure}
    \centering
    \includegraphics{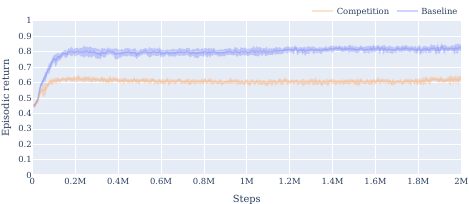}
    \caption{Mean and 99\% confidence interval of episodic return during training of our GNN policy based on results from 10 different random seeds. The results from training against the baseline agents and training against the competition agents are plotted.}
    \label{fig:training_random}
\end{figure}

\subsubsection{Baseline Opponents}\label{sec:result_baseline}
\begin{figure}
    \centering
     \begin{subfigure}{0.27\textwidth}
         \centering
         \includegraphics{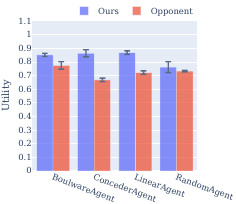}
         \caption{}\label{fig:random1}
     \end{subfigure}
     \begin{subfigure}{0.72\textwidth}
         \centering
         \includegraphics{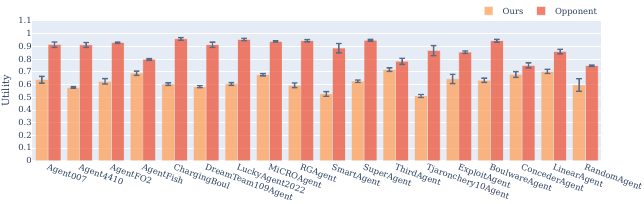}
         \caption{}\label{fig:random2}
     \end{subfigure}
    \caption{Evaluation results of our GNN-based policy on randomly generated negotiation problem both against the set of baseline opponents (left) and against the full set of opponents (right). Results are obtained by evaluating each trained policy for 1000 negotiation games against the set of agents. Mean and 99\% confidence interval are plotted based on 10 training iterations.}
    \label{fig:random}
\end{figure}

We first train and evaluate against the set of baseline agents as described in \autoref{tab:opp}. We train our method for 2\,000\,000 steps on 10 random seeds. The learning curve is plotted in \autoref{fig:training_random}. Results are again obtained by running 1000 negotiation sessions against the set of baseline opponents, but this time, all negotiation problems are randomly generated and are never seen before. We note that the observation and action space sizes are constantly changing. Results are plotted in \autoref{fig:random1}.

As seen in \autoref{fig:random1}, our method performs well against all baseline agents while negotiating on various structured negotiation problems it has never seen before. It is promising that an end-to-end learned GNN-based policy appears to generalise over such different problems.

\subsubsection{Competition Opponents}\label{sec:result_all}
We now repeat the experiment from \autoref{sec:result_baseline}, but increase the set of agents we negotiate against. More specifically, we add the agents of the 2022 edition of the Automated Negotiation Agents Competition (ANAC)\footnote{\url{https://web.tuat.ac.jp/~katfuji/ANAC2022/}}. The learning curve and results are plotted in \autoref{fig:training_random} and \autoref{fig:random2}, respectively.

The results show much lower performance against all opponents, including those outperformed in \autoref{sec:result_baseline}. Our current method of encoding the observations and design of the policy likely leads to limited capabilities of learning opponent characteristics. Past work has shown that adapting to opponents is important to improve performance~\citep{ilany_algorithm_2016,sengupta_autonomous_2021,renting_automated_2022}, which is currently impossible. However, this goes beyond the core contribution of this work, which is about handling different-sized negotiation problems in end-to-end RL methods. We discuss potential solutions in \autoref{sec:conclusion}.

%% file: sections/5_conclusion.tex
We developed an end-to-end RL method for training negotiation agents capable of handling differently structured negotiation problems. We showed that our method performs as well as a recent end-to-end method that is not transferrable beyond a single fixed negotiation problem. We see the latter as a serious restriction since, in real-world applications, it would be extremely unlikely to encounter the exact same negotiation problem more than once.

In our work presented here, for the first time, we have demonstrated how the difficulty of dealing with changing negotiation problems in end-to-end RL methods can be overcome. 
Specifically, we have shown how an agent can learn to negotiate on diverse negotiation problems in such a way that performance generalises to never-before-seen negotiation problems. Our method is conceptually simple compared to previous work on reinforcement learning in negotiation agents. Our agent performs well against strong baseline negotiation strategies, but leaves room for improvement when negotiating against a broad set of highly competitive agents. 

Our approach is based on encoding the stream of observations received by our agent into a graph whose node features are designed to capture historical statistics about the episode. This manual feature design likely leads to information loss and goes against the end-to-end aim of our approach. For example, the expressiveness of history is limited as the graph only encodes the last offer and frequency of offers. This likely also causes limited adaptivity to a broad set of opponent strategies, which in turn causes the low performance observed in \autoref{sec:result_all}.

We note that,  due to the competition setup of ANAC,  competitive agents often play a game of chicken. Performing well against such strategies means that a policy must also learn this game of chicken. This can be challenging for RL, due to exploration problems, as it must learn a long sequence of relatively meaningless actions before having a chance to select a good action. We could attempt to improve upon this, but it might be more beneficial to prioritize mitigating this game of chicken behaviour, as it is inefficient and (arguably) undesirable.

The negotiation problems we generated have additive utility functions and a relatively small outcome space, as is quite typical for benchmarks used in automated negotiation research. 
Real-world negotiation problems, however, can have huge outcome spaces~\citep{de_jonge_nb3_2015}. Our designed policy can be applied to larger problems without increasing the trainable parameters, and the effects on the performance of doing this should be investigated in future work.

Further promising avenues for future work include extending end-to-end policies with additional components that, e.g., learn opponent representations based on the history of observations in the current or previous encounter. Changing a negotiation strategy based on the opponent characteristics has been shown previously to improve performance~\citep{ilany_algorithm_2016,sengupta_autonomous_2021,renting_automated_2022}, but is likely difficult to learn through our current policy design. Furthermore, improving our method to handle continuous objectives would eliminate the necessity of discretizing them.

Overall, we believe that in this work, we have taken a substantial step towards the effective use of end-to-end RL for the challenging and important problem of training negotiation agents whose performance generalises to new negotiation problems and opens numerous exciting avenues for future research in this area.

%% file: sections/A_hyperparameters.tex
\begin{table}[h]
    \centering
    \begin{tabular}{lc}
        \toprule
        \textbf{Parameter} & \textbf{Value} \\
        \midrule
        total timesteps & $2 \cdot 10^6$\\
        batch size & 6000 \\
        mini batch size & 300 \\
        policy update epochs & 30 \\
        entropy coefficient & 0.001 \\
        discount factor $\gamma$ & 1 \\
        value function coefficient & 1 \\
        GAE $\lambda$ & 0.95 \\
        \# GAT layers & 4 \\
        \# GAT attention heads & 4 \\
        hidden representation size & 256 \\
        Adam learning rate & $3 \cdot 10^{-4}$ \\
        Learning rate annealing & True \\
        activation functions & ReLU \\
        \bottomrule
    \end{tabular}
    \caption{Hyperparameter settings}
    \label{tab:hp}
\end{table}